\documentclass[prd,twocolumn,nofootinbib,notitlepage,aps,tightenlines,
preprintnumbers,amsmath,amssymb,amsfonts,showpacs,superscriptaddress]{revtex4-1}

\usepackage[hyperindex,breaklinks]{hyperref}
\usepackage{graphicx,epstopdf}
\usepackage{mathtools}
\usepackage{natbib,bm}
\usepackage[usenames,dvipsnames]{color}
\usepackage{slashed}    
\usepackage{hyperref}
\usepackage{multirow}

\def\be{\begin{equation}}
\def\ee{\end{equation}}
\def\ba{\begin{eqnarray}}
\def\ea{\end{eqnarray}}
\def\ge{\mathrel{\raise.3ex\hbox{$>$\kern-.75em\lower1ex\hbox{$\sim$}}}}
\def\la{\mathrel{\raise.3ex\hbox{$<$\kern-.75em\lower1ex\hbox{$\sim$}}}}

\def\simgt{\mathrel{\raise.3ex\hbox{$>$\kern-.75em\lower1ex\hbox{$\sim$}}}}
\def\simlt{\mathrel{\raise.3ex\hbox{$<$\kern-.75em\lower1ex\hbox{$\sim$}}}}

\newcommand{\nc}{\newcommand}

\nc{\gone}{\bar g_{\pi NN}^{(1)}}
\nc{\gzero}{\bar g_{\pi NN}^{(0)}}
\nc{\al}{\alpha}
\nc{\ga}{\gamma}
\nc{\de}{\delta}
\nc{\ep}{\epsilon}
\nc{\ze}{\zeta}
\nc{\et}{\eta}
\nc{\ka}{\kappa}
\nc{\rh}{\rho}
\nc{\si}{\sigma}
\nc{\ta}{\tau}
\nc{\up}{\upsilon}
\nc{\ph}{\phi}
\nc{\ch}{\chi}
\nc{\ps}{\psi}
\nc{\om}{\omega}
\nc{\Ga}{\Gamma}
\nc{\De}{\Delta}
\nc{\La}{\Lambda}
\nc{\Si}{\Sigma}
\nc{\Up}{\Upsilon}
\nc{\Ph}{\Phi}
\nc{\Ps}{\Psi}
\nc{\Om}{\Omega}
\nc{\ptl}{\partial}
\nc{\del}{\nabla}
\nc{\ov}{\overline}
\nc{\newcaption}[1]{\centerline{\parbox{15cm}{\caption{#1}}}}
\nc{\us}{U(1)$_S$}

\def\beq{\begin{equation}}
\def\eeq{\end{equation}}
\def\bmat{\begin{displaymath}}
\def\emat{\end{displaymath}}
\def\bear{\begin{eqnarray}}
\def\eear{\end{eqnarray}}
\def\ba{\begin{eqnarray}}
\def\ea{\end{eqnarray}}
\def\bery{\begin{array}}
\def\ery{\end{array}}
\def\bit{\begin{itemize}}
\def\eit{\end{itemize}}
\def\ben{\begin{enumerate}}
\def\een{\end{enumerate}}
\def\btab{\begin{tabular}}
\def\etab{\end{tabular}}
\def\btbl{\begin{table}}
\def\etbl{\end{table}}
\def\bfig{\begin{figure}[htb]}
\def\efig{\end{figure}}
\def\bpic{\begin{picture}}
\def\epic{\end{picture}}

\def\ga{\mathrel{\raise.3ex\hbox{$>$\kern-.75em\lower1ex\hbox{$\sim$}}}}
\def\la{\mathrel{\raise.3ex\hbox{$<$\kern-.75em\lower1ex\hbox{$\sim$}}}}
\def\gappeq{\mathrel{\rlap {\raise.5ex\hbox{$>$}}
{\lower.5ex\hbox{$\sim$}}}}
\def\lappeq{\mathrel{\rlap{\raise.5ex\hbox{$<$}}
{\lower.5ex\hbox{$\sim$}}}}

\def\gyr{{\rm \, G\kern-0.125em yr}}
\def\mev{{\rm \, Me\kern-0.125em V}}
\def\gev{{\rm \, Ge\kern-0.125em V}}
\def\tev{{\rm \, Te\kern-0.125em V}}

\begin{document}

\title{LSND Constraints on the Higgs Portal}

\author{Saeid Foroughi-Abari}
\affiliation{Department of Physics and Astronomy, University of Victoria, 
Victoria, BC V8P 5C2, Canada}

\author{Adam Ritz}
\affiliation{Department of Physics and Astronomy, University of Victoria, 
Victoria, BC V8P 5C2, Canada}

\date{April 2020}

\begin{abstract}
\noindent 

High-luminosity fixed target experiments provide impressive sensitivity to new light weakly coupled degrees of freedom. We revisit the minimal case of a scalar singlet $S$ coupled to the Standard Model through the Higgs portal, that decays visibly to leptons for scalar masses below the di-pion threshold. The dataset from the LSND experiment is found to impose the leading constraints within two mass windows between $m_S \sim 100$ and 350 MeV. In the process, we analyze a number of scalar production channels in the target, finding that proton bremsstrahlung provides the dominant channel at LSND beam energies.

\end{abstract}
\maketitle

\section{Introduction}

The empirical evidence for physics beyond the Standard Model (SM), notably for dark matter and neutrino mass, may point to the presence of a more complex hidden (or dark) sector \cite{pospelov2008,Batell:2009yf,Essig:2009nc,Reece:2009un,Freytsis:2009bh,Batell:2009jf,Freytsis:2009ct,Essig:2010xa,Essig:2010gu,McDonald:2010fe,Williams:2011qb,Abrahamyan:2011gv,Archilli:2011zc,Lees:2012ra,Davoudiasl:2012ag,Kahn:2012br,Andreas:2012mt}. The defining feature of such scenarios is the presence of degrees of freedom which are weakly coupled to the SM, and therefore may be light relative to the weak scale. As a result, dark sectors with light degrees of freedom can be probed with a variety of experiments at the luminosity frontier, including proton \cite{Batell:2009di,deNiverville:2011it,deNiverville:2012ij,Kahn:2014sra,Adams:2013qkq,Soper:2014ska,Dobrescu:2014ita,Coloma:2015pih,dNCPR,MB1,MB2,Alpigiani:2018fgd,Ariga:2018pin} and electron \cite{Bjorken:2009mm,Izaguirre:2013uxa,Diamond:2013oda,Izaguirre:2014dua,Batell:2014mga,Lees:2017lec,Berlin:2018bsc,NA64:2019imj} fixed target facilities. This framework has been explored in great detail over the past decade (see e.g. \cite{DS16,CV17,PBC}).

From an effective field theory perspective, classifying the interactions of new neutral states with the Standard Model (SM) according to their dimensionality, there are only three relevant or marginal `portal' operators that are not suppressed by a new energy scale. The Higgs, vector and neutrino portals therefore comprise the leading couplings of the SM to a hidden or dark sector. Motivated in part by the phenomenology of light dark matter (DM), much theoretical and experimental effort has recently been focussed on these portals  \cite{DS16,CV17}.

In this paper, we will consider the minimal Higgs portal \cite{PW}, the unique relevant operator that can couple the SM model to a dark sector,
\be
 {\cal L}_{SH} \supset - A S H^\dagger H,
\ee
where $S$ is a new scalar singlet, $H$ is the SM Higgs doublet, and $A$ is a dimensional portal coupling. Along with being one of the few renormalizable portal couplings to a dark sector, and a potential force mediator for thermal relic models of light dark matter, this interaction is of intrinsic interest as an extension of the SM Higgs sector. 

The strongest existing constraints on the Higgs portal, in the low mass range where Br$(S\rightarrow l^+l^-) \sim 1$, arise from searches for leptonic decays at the CHARM fixed target experiment at CERN \cite{Bergsma:1985qz,Clarke:2013aya,Winkler:2018qyg}, and analysis of $K^+ \rightarrow \pi^+ S$ signatures at the Brookhaven E949 experiment, with $S$ escaping the detector before decaying and thus being counted as missing energy in the search for $K^+ \rightarrow \pi^+ \nu\bar{\nu}$ \cite{Artamonov:2009sz,Clarke:2013aya,Winkler:2018qyg}. The latter constraint is the most stringent, except in an $S$ mass range relatively close to $m_\pi$ where significant backgrounds limit the reach of E949. For higher $S$ masses, a range of accelerator and meson decay constraints apply to the Higgs portal \cite{Batell:2009jf,Aaij:2016qsm,Clarke:2013aya,Pospelov:2017kep,FP17,Winkler:2018qyg,BB1,BB2,BB3}, while for smaller couplings, constraints from supernova cooling also apply \cite{Krnjaic:2015mbs,Dev:2020eam}.

In this paper, we revisit the limits on the Higgs portal in the low mass $m_S <$ 350 MeV region, by studying the sensitivity of the LSND experiment, that is known to provide important constraints on the dark photon \cite{Batell:2009di,deNiverville:2011it}. In particular, by analyzing a range of production channels in the interaction of the 800 MeV proton beam with the target, particularly proton bremsstrahlung \cite{BB2}, we find that existing LSND analyses with final state electrons and muons already impose the leading constraint on the Higgs portal in two mass windows between 100 and 350~MeV. Our final results are presented in Fig.~1, where we show LSND exclusions compared to existing limits e.g. from CHARM  and E949, and recent projections for sensitivity at the Fermilab SBN facility \cite{Batell:2019nwo}. For comparison, we also show the 1 and 2$\si$ preferred regions if decays of $K_L$ via $S$ were to explain the recent KOTO anomaly \cite{Ahn:2018mvc,KOTO,Egana-Ugrinovic:2019wzj,Dev:2019hho,Liu:2020qgx}.

The rest of this paper is organized as follows. In the next Section we define the Higgs portal, and summarize some of the relevant couplings and decays rates.
In Section~3, the production of light scalars at LSND is discussed in some detail, and in Section~4 we present the sensitivity reach due to light scalars decaying to electrons and muons in the detector. Section~5 contains our concluding remarks.

\begin{figure}[t]
\vspace*{-0.5cm}
 \centerline{\hspace*{0.5cm}
 \includegraphics[width=0.53\textwidth]{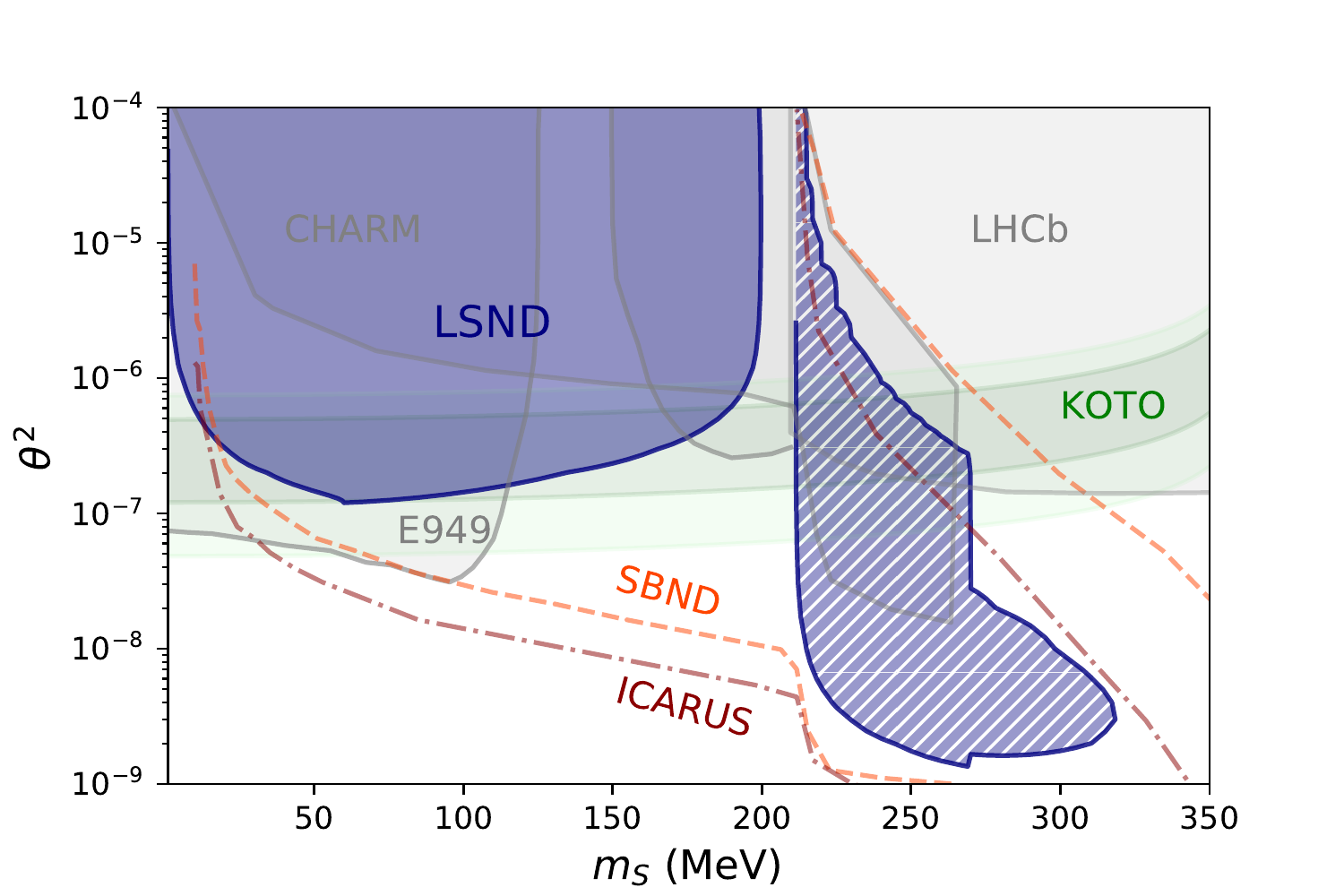}  
 }
 \caption{\footnotesize A summary of the sensitivity limits determined in this work for scalar $S$ decays to electrons (solid) and muons (hatched) at LSND, shown in the plane of the coupling $\theta^{2}\simeq (Av/m_h^2)^2$ versus dark scalar mass $m_{S}$. Exclusions from other sources (in gray) including LHCb \cite{Aaij:2016qsm}, E949 $ K \rightarrow \pi+ invisible$ \cite{Artamonov:2009sz,Clarke:2013aya,Winkler:2018qyg}, and CHARM $S\rightarrow e^{+}e^{-},\; \mu^{+}\mu^{-}$  \cite{Bergsma:1985qz,Clarke:2013aya,Winkler:2018qyg} are shown. The 1 and 2$\si$ preferred contours to explain the KOTO anomaly in $K_L$ decays \cite{Ahn:2018mvc,KOTO,Egana-Ugrinovic:2019wzj,Dev:2019hho}, and the sensitivity projections for the on-axis SBND (orange) and off-axis ICARUS (purple) experiments at Fermilab \cite{Batell:2019nwo}, are also shown for comparison (see the text for further details). }
 \label{sensitivity}
\end{figure}

\section{Higgs portal}
\label{sec:hp}

We extend the SM by adding a scalar singlet $S$, for which the leading relevant or marginal couplings to the Higgs doublet $H$ comprise the Higgs portal \cite{PW},
\be
 {\cal L} \supset - (AS + \lambda S^2) H^\dagger H.
\ee
After electroweak symmetry breaking, and re-diagonalizing by shifting the physical Higgs field as $h \rightarrow h + \theta S$ to remove $hS$ mixing,  the induced linear couplings take the form
\begin{align}
 {\cal L} &\supset 
   - \frac{\theta}{v} S \left( \sum_f m_f \bar{f} f + m_Z^2 Z^\mu Z_\mu + 2m_W^2 W^+_\mu W^{\mu+}\right), \nonumber
\end{align}
where the mixing angle $\theta \simeq Av/m_h^2 \ll 1$ for the parameters of interest in this paper. 

Integrating out the electroweak-scale degrees of freedom induces further couplings of $S$ to light hadronic states. For the sub-GeV mass range of interest here, the relevant interactions take the form,
\begin{align}
 {\cal L} &\supset 
   - \theta S \left(  \frac{m_e}{v} \bar{e} e + g_{S\gamma\gamma} F_{\mu\nu} F^{\mu\nu} + g_{SNN} \bar{N} N + \cdots \right).\nonumber
\end{align}
Well-known 1-loop triangle diagrams generate the effective diphoton coupling \cite{Djouadi:2005gi},
\be \label{diphoton}
 g_{S\gamma\gamma} = \frac{\al}{8\pi v} F_\gamma (m_S),
\ee
where $F_\gamma (m_S \ll {\rm GeV} )\sim {\cal O}(1)$ is a loop function \cite{Djouadi:2005gi,Fradette:2018hhl,BB2}. The coupling to nucleons can in turn be obtained through the use of low energy theorems \cite{BB2} (see also \cite{Alarcon:2011zs,Alarcon:2012nr}),
\begin{align}
 g_{SNN} &\simeq \frac{2}{9} \frac{m_N}{v} \left(1+\frac{7}{2} \sum_{q=u,d,s} \frac{m_q}{m_N} \langle N| \bar{q} q |N \rangle \right) \nonumber\\
  &\sim 1.2 \times 10^{-3}.
\end{align}
In principle this coupling should be extended to a form-factor, but for the kinematic regime of interest in this paper, there is no significant impact from hadronic scalar resonances, and the assumption that $g_{SNN}$ is a constant will be sufficient.

In analyzing the fixed target detection signatures of $S$ decays, we will also require the leptonic decay width of $S$, which is given by \cite{Batell:2009jf},
\be
 \Ga(S\rightarrow l^+l^-) = \theta^2 \frac{m_l^2 m_S}{8\pi v^2} \left( 1- \frac{4m_l^2}{m_S^2} \right)^{3/2}.
\ee
We have Br$(S\rightarrow e^+e^-)\simeq 1$ for $2m_e < m_S < 2 m_\mu$, which is the dominant decay channel over much of the mass range of interest here, while Br$(S\rightarrow \mu^+\mu^-)\simeq 1$ for $2m_\mu < m_S < 2 m_\pi$. Just above the pion threshold, Br$(S\rightarrow \mu^+\mu^-)\simeq 0.15-0.2$ \cite{FP17}, which will also be relevant below.

\section{Light scalar production at LSND}
\label{production}

The LSND experiment comprises an 800 MeV proton beam impacting a thick target, that was either water or a high $Z$ metal at various stages of the experimental program. Over its lifetime LSND accumulated one of the largest proton on target (POT) datasets of any fixed target experiment, with over $10^{23}$ POT in total \cite{Athanassopoulos:1997er, Aguilar:2001ty}. The relevance of LSND for Higgs portal phenomenology was briefly addressed in \cite{Pospelov:2017kep}. In this section, we will revisit the production rate of scalars for $m_S < m_\pi$ from a variety of channels. 

Before we examine specific production modes, it is useful to compare this case to the scenario with a dark photon $A'_\mu$ kinetically mixed with the photon via the interaction $\frac{\ep}{2} F^{\mu\nu} F'_{\mu\nu}$. This induces a low energy coupling of $A'$ to the electromagnetic current with strength $e\ep$, with $\ep$ the kinetic mixing parameter. The leading production mode at LSND for low mass dark photons is pseudoscalar meson decay, e.g. Br$(\pi^0 \rightarrow A'\gamma) \sim \ep^2$. Thus, for sufficiently light dark photons, we can estimate the number of dark photons produced as $N^{(\pi)}_{A'} \sim  \ep^2 N_\pi$. The large pion (and eta) production rate, combined with the large radiative branching for pseudoscalar mesons makes this channel by far the most efficient. For scalars coupled through the Higgs portal, the situation is somewhat different, as the only mesons with substantial scalar branching rates are kaons and $B$ mesons, which are not kinematically accessible at LSND. We find instead that the dominant production mode in this case is proton bremsstrahlung, $p+N \rightarrow X + S$, with $X$ the inclusive hadronic final state and $N_S \sim 0.5 \theta^2 g_{SNN}^2 N_\pi \sim 10^{-6} \theta^2 N_\pi$, which is substantially lower than the dark photon production rate due to the reduced scalar coupling to hadrons. In the rest of this section, we will discuss this production mode in more detail, along with further production channels via $\De$ decay and the Primakov process for comparison.


\begin{figure}[t]
 \centerline{
 \includegraphics[width=0.4\textwidth]{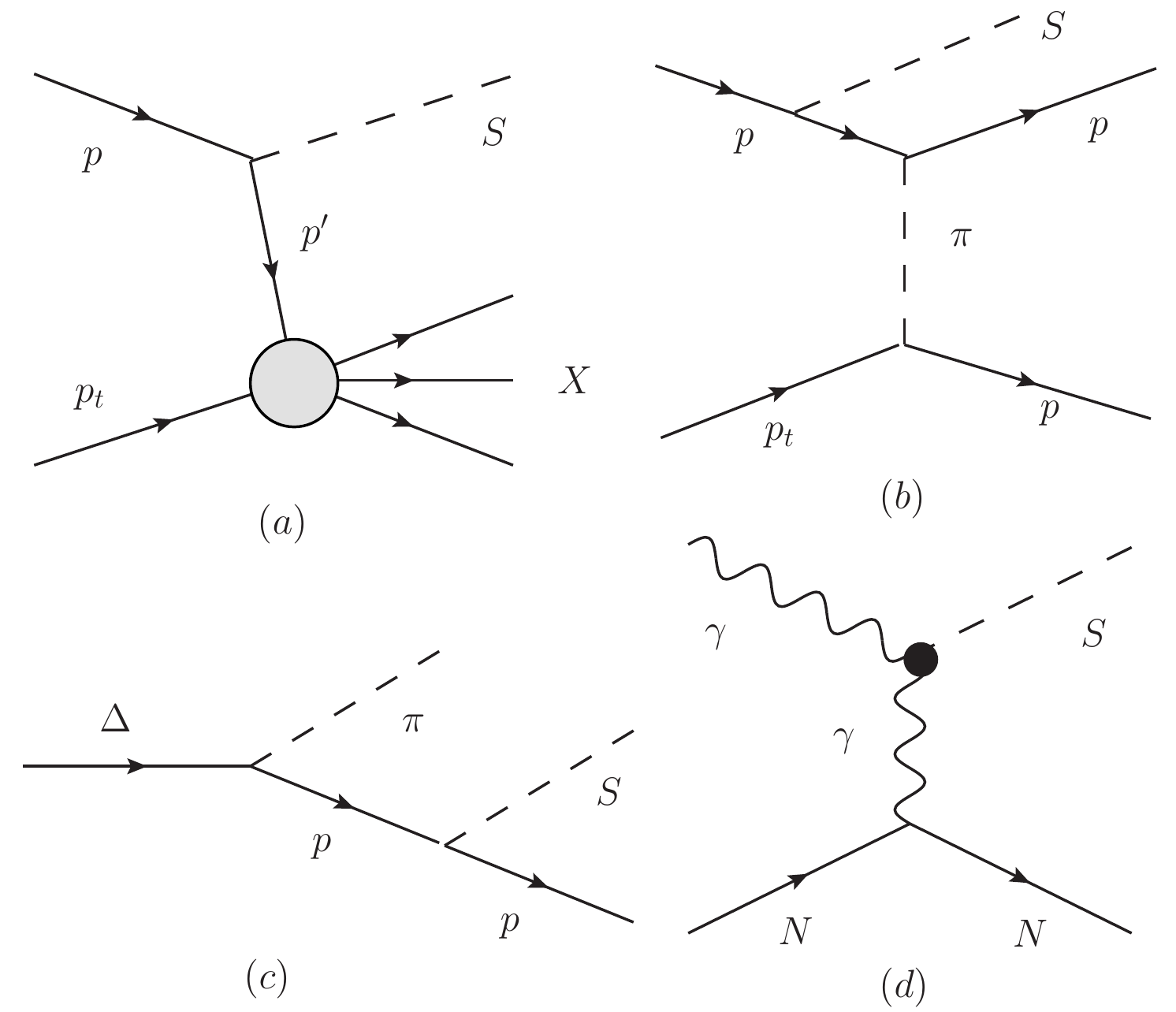}  
 }
 \caption{\footnotesize Production channels for the scalar S at LSND via (a) proton bremsstrahlung splitting function, (b) proton bremsstrahlung through one-pion exchange, (c) $\Delta$ decay and (d) the Primakov process.}
 \label{Feynman}
\end{figure}

\subsection{Proton bremsstrahlung}
Scalars can be produced through the $ SNN $ vertex via the proton-proton bremsstrahlung process $ p+p\rightarrow S+X$, where we focus on $pp$ scattering due to its resonantly enhanced rate, proceeding via the $\De^{++}$ intermediate state. At LSND beam energies, the beam protons are only moderately relativistic, and thus we will utilize two different procedures for the calculation adapted respectively to either sub-relativistic or highly relativistic beams. Comparing the scalar production rate using both techniques at LSND will allow for an assessment of the precision of the rate calculation.

{\it Splitting function:-} We will first follow the approach of Altarelli-Parisi and formulate the bremsstrahlung calculation in quantum mechanical perturbation theory, as recently discussed in this context in \cite{BB2}. Since the beam protons are not ultra-relativistic at LSND, this is an extension of the conventional Weizsacker-Williams (WW) approximation in which the beam protons are often considered in the infinite-momentum frame. Nonetheless, we find that the kinematic range  at LSND will still allow us to approximate the required rate in terms of the proton-proton cross-section and a calculable sub-process \cite{PhysRevD.12.187,Altarelli:1977zs,Liu:2016mqv,Boiarska:2019jym}. In this formalism, all states are on-shell and while 3-momentum is conserved, energy is not automatically conserved at each vertex, but only after summing all contributions. 

The relevant diagram for the process is shown in Fig.~\ref{Feynman}(a). We denote the momentum of the incoming proton and emitted $S$ in the target rest frame by $ p_{p}^{\mu}{=}(E_{p},\vec{0},p_{p}) $ and $p^{\mu}_{S}{=}(E_{S},\vec{p}_{T},zp_{p})$, with $E^2_{S}{=}z^{2}p_{p}^{2}{+}p_{T}^{2}{+}m_{S}^{2}$. The momentum of the intermediate proton is denoted $ p_{p^{\prime}}^{\mu}{=}(E_{p^{\prime}},-\vec{p}_{T},(1{-}z)p_{p})$, with $E^2_{p^{\prime}}{=}(1{-}z)^{2}p_{p}^{2}{+}p_{T}^{2}{+}m_{p}^{2}$, where $ p_{T} $ is the $ S $ transverse momentum with respect to the beam, and $ z $ is the fraction of longitudinal momentum carried by $ S $.

\begin{figure}[t]
 \centerline{\hspace*{0.5cm}
 \includegraphics[width=0.53\textwidth]{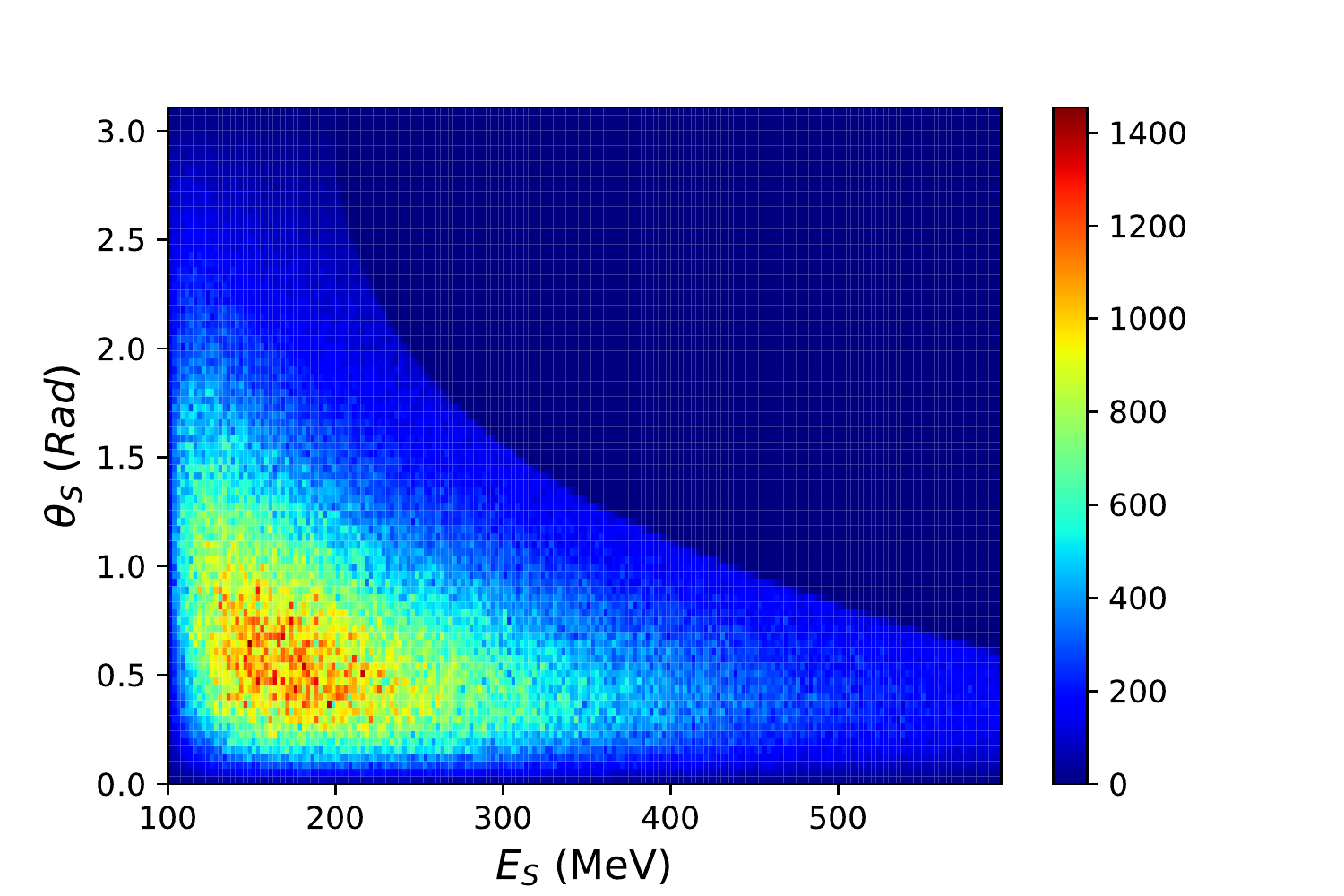}  
 }
 \caption{\footnotesize Energy-angle distribution of scalars with $m_{S}=100$ MeV produced via the proton bremsstrahlung channel at the LSND beam energy of $0.8$ GeV. This distribution uses the assumptions discussed in the text, and has an arbitrary overall normalization with the colour bar indicating the relative frequency.}
 \label{lab_Dist}
\end{figure}

\begin{figure}[t]
 \centerline{
 \includegraphics[width=0.45\textwidth]{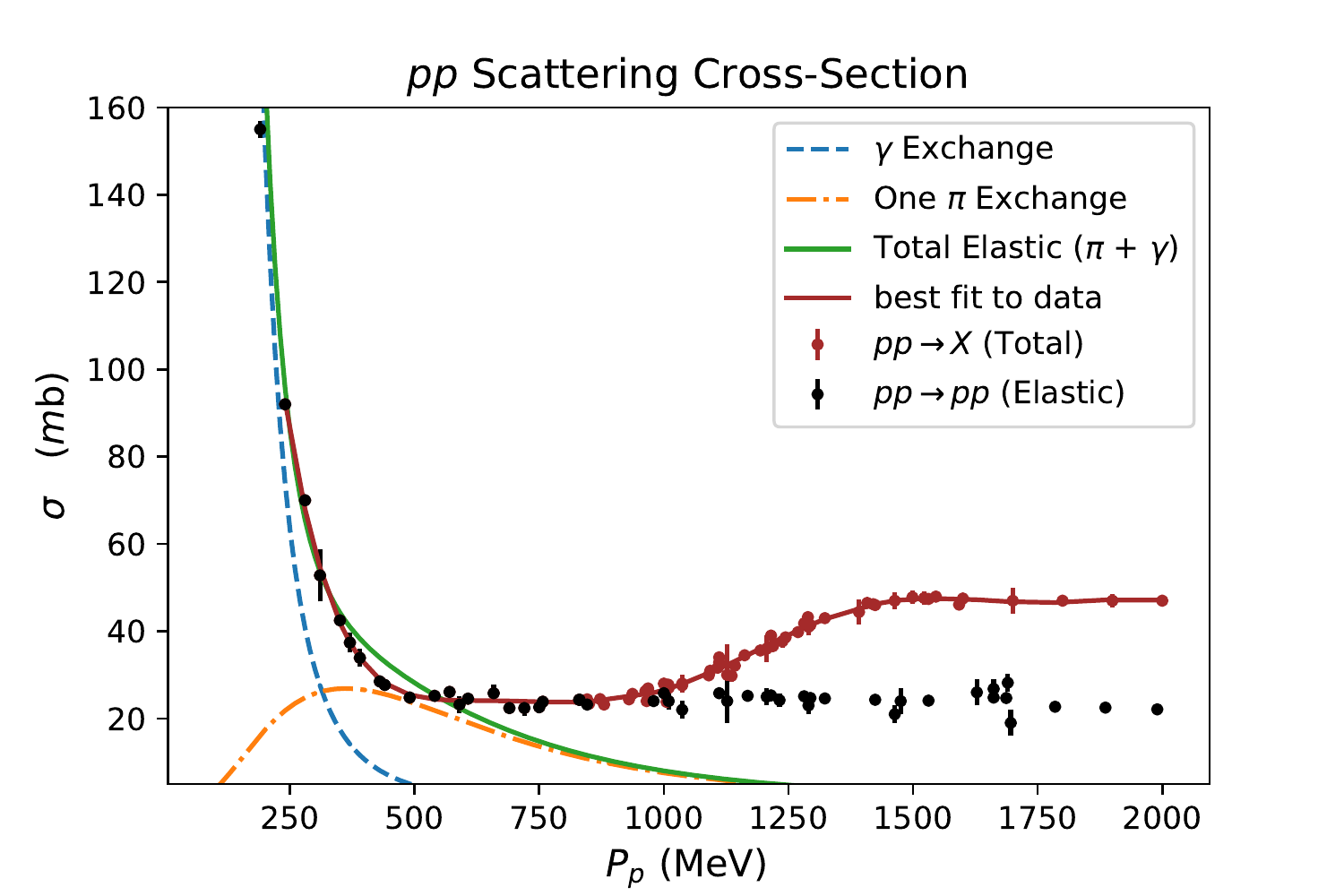}  
 }
 \caption{\footnotesize The $pp$ - scattering cross-section as a function of the proton beam 3-momentum (MeV). The curves denote the contributions from one pion exchange (dash-dotted line), photon exchange (dashed line), and the total cross section (solid line), compared with data for elastic and inelastic scattering from the Particle Data Group \cite{PDBook}. For the electromagnetic component there is a cut on the forward/backward angle of the scattered proton in the lab frame of $2^\circ$.}
 \label{fig1}
\end{figure}

The second order contribution to the matrix element, generically of the form $V_{fj} V_{ji}/(E_f - E_i)$ for a perturbation $V$, has two possible time orderings in this case for the process $p{+}p_t{\rightarrow}S{+}X$ exchanging the intermediate state $ p^{\prime}$. The two amplitudes can be written as \cite{Altarelli:1977zs,BB2}
\begin{align}\label{twomatrix}
& \mathcal{M}^{\rm emit} = \frac{\mathcal{M}_{p\rightarrow p^{\prime}S}\mathcal{M}_{pp^{\prime}\rightarrow X}}{2E_{p^{\prime}}(E_{p}-E_{S}-E_{p^{\prime}})} \\  
& \mathcal{M}^{\rm absorb} =\frac{\mathcal{M}_{p\rightarrow p^{\prime}X}\mathcal{M}_{pp^{\prime}\rightarrow S}}{2E_{p^{\prime}}(E_{S}-E_{p}-E_{p^{\prime}})} 
\end{align}
where the intermediate proton's $3$-momentum is fixed by $ \vec{p}_{p^{\prime}}{=}\vec{p}_{p}{-}\vec{p}_{S} $, while the energy is not automatically conserved at the $ pp^{\prime}S $ vertex.
Denoting the energy denominators as $\Delta E_{\rm emit}{=}E_{p^{\prime}}{+}E_{S}{-}E_{p}$ and $ \Delta E_{\rm absorb}{=}E_{p^{\prime}}{-}E_{S}{+}E_{p}$, then under the condition
\be \label{assumptions}
\Delta E_{\rm emit}\ll \Delta E_{\rm absorb},
\ee
we can neglect the matrix element $\mathcal{M}^{\rm absorb}$. This can be interpreted as the dominant contribution coming from initial state radiation. We have verified that this condition is satisfied to a few percent for LSND kinematics. Imposing a second condition,
\be 
 \Delta E_{\rm emit}\ll m_{p}, \label{assumption2}
 \ee 
it is possible to write the differential cross section of the process $p{+}p_t{\rightarrow}S{+}X $ in the approximate form \cite{Boiarska:2019jym},
\be
\frac{d\sigma_{pp_t\rightarrow SX}}{dz dp_T^2} \approx P^{\rm split}_{S}(z,p_{T})\sigma_{pp}(s^{\prime}),
\label{Brem}
\ee
where $\sigma_{pp}$ is the total proton-proton scattering cross section, which varies between $\sim 30{-}45$mb over the relevant energy range (see Fig.~\ref{fig1}) \cite{PhysRevD.98.030001}, with $s^{\prime}{=}2m_p(E_{p}{-}E_{S}{+}m_p)$ the center of mass energy. Denoting the momentum transfer as $ q^{\mu} =(E_{p}{-}E_{S},\vec{p}_{p}{-}\vec{p}_{S}) $, the differential splitting probability of the proton to emit a scalar $P^{\rm split}_{S}$ can be represented in the form,
\begin{align}
P^{\rm split}_{S}(z,p_{T}) =& \frac{\theta^{2} g_{SNN}^{2}}{16 \pi^{2}} \frac{(E_{p}E_{p^{\prime}}{-}(1{-}z)p_{p}^{2}{+}m_{p}^{2})}{E_{S}(q^{2}{-}m_{p}^{2})^{2}} \times \nonumber \\
& \sqrt{(1{-}z)^{2}p_{p}^{2}+p_{T}^{2}}\bigg(1+\frac{E_{p}{-}E_{S}}{E_{p^{\prime}}}\bigg)^{2} .
\end{align}
The integration range for $p_{T}$ and $z$ is determined by the kinematic conditions (\ref{assumptions},\ref{assumption2}), where we require the kinematic variable on the left of each inequality to be at most 10\% of the right hand side. The conditions are satisfied for $z\in [0, 0.5]$ and $p_{T}<300$ MeV at LSND. The resulting distribution of scalars is shown in Fig.~\ref{lab_Dist}, which we see reaches above $E_S \sim 300$~MeV.

{\it One pion exchange:-} We will now consider a complementary approach, modelling proton-proton scattering via one pion exchange, which is expected to provide the dominant hadronic (as opposed to electromagnetic) contribution to bremsstrahlung at sub-relativistic beam energies. Using ${\cal L} = g_{\pi NN} \bar{N} \gamma_5 \tau \cdot \pi N$, with $g^2_{\pi NN} /(4\pi) \approx 13.5$, we first verify that the tree-level one pion exchange contribution to $pp$ elastic scattering does provide a relatively good fit, after accounting for the electromagnetic component, as shown in Fig.~\ref{fig1}. We utilize a dipole form for the pion-nucleon form factor $\sim 1/(1+Q^2/m_A^2)^2$ where $m_A \sim 1\,{\rm GeV}$ is the axial mass \cite{Schindler:2006it,Megias:2019qdv},
and similarly the proton electromagnetic form factor $F_1(Q^2) \sim 1/(1+Q^2/(0.71 {\rm GeV})^2)^2$. The contribution from one pion exchange is significant in a narrow energy range, and it is known that additional processes, such as two pion exchange, become important for beam momenta above 600-700 MeV \cite{PhysRevC.77.014003,PhysRev.176.1762}. Retaining just the one pion exchange contribution will nonetheless be sufficient in our case, as we are interested in the ratio of two- to three-body final states, in which the overall normalization of the $pp$ cross section drops out as for the splitting function calculation above.  Note that above a beam momentum of about a GeV, the inelastic channel via the $\De$-resonance contributes at a comparable level to elastic scattering, but is not accounted for in this approximation. 

We now compute the rate for initial state radiation of $S$, $pp\rightarrow ppS$ via one pion exchange, according to Fig.~\ref{Feynman}(b). For the analysis below, we use the full tree-level calculation of the 2-body and 3-body final states. However, we can gain some intuition in the limit where Mandelstam $s \gg m_S^2$, where the cross section takes the form  $\sigma_{pp\rightarrow ppS} = \frac{g_{SNN}^2}{8\pi^2} \sigma_{pp} f(m_S^2)$, with $f(m_S^2) \propto \log^2 \frac{m_S^2}{s} + \cdots$ exhibiting the Sudakov double logarithim. For the finite-$m_S$ kinematics of interest here, there are no sizeable IR/collinear effects, and so we will not need to include the corresponding loop contribution that is relevant in the $m_S \rightarrow 0$ limit.

\begin{figure}[t]
 \centerline{
 \includegraphics[width=0.45\textwidth]{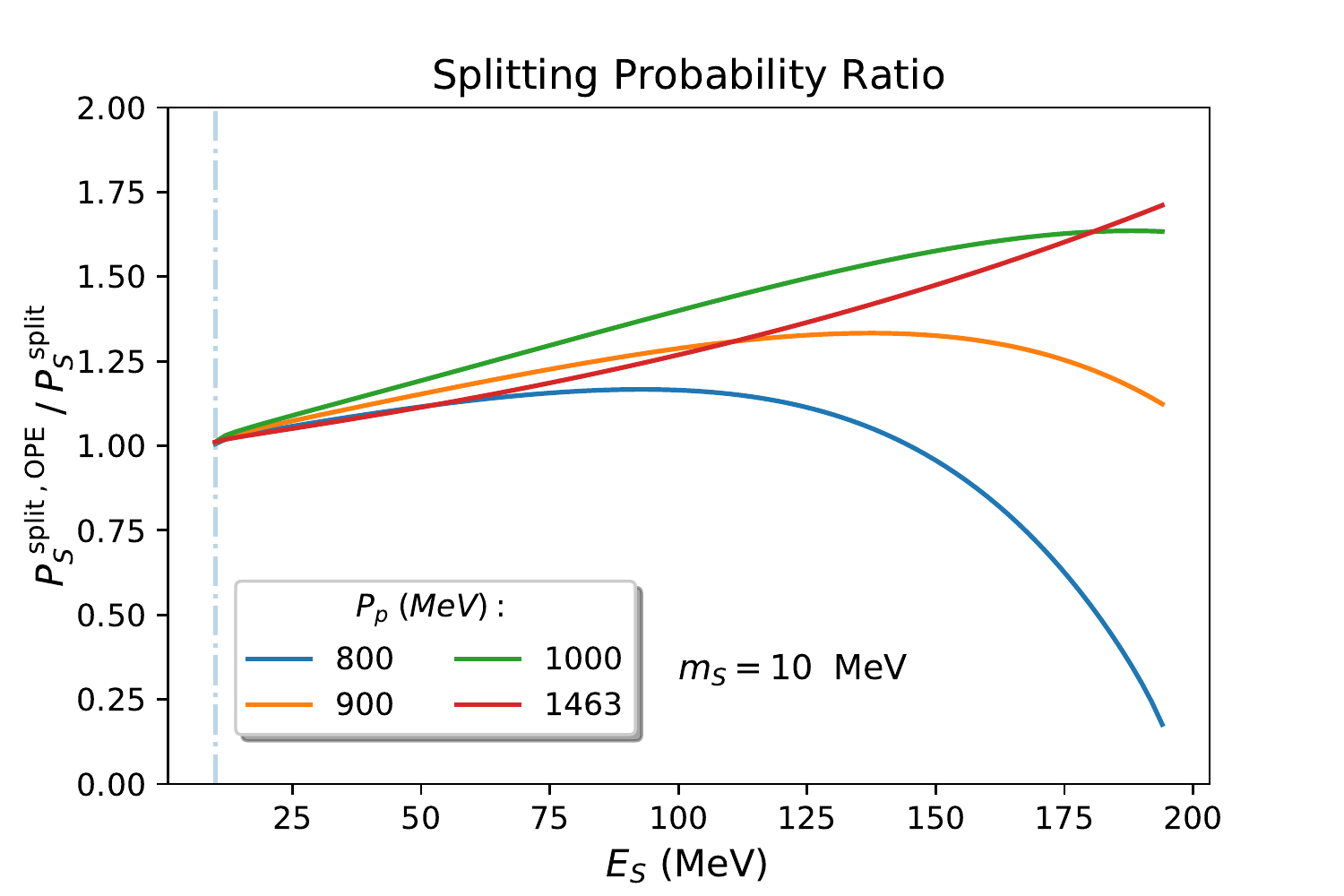}  
 }
 \caption{\footnotesize The ratio of the splitting probability of the initial state proton to emit a scalar calculated using the two techniques as a function of scalar energy. The LSND beam momentum corresponds to $P_p = 1463\,$MeV.}
 \label{fig2}
\end{figure}

To compare with the splitting function calculation above, we define the differential splitting probability of the proton to emit a scalar via one pion exchange in the form,
\be
P_{S}^{\mathrm{split, OPE}}(E_S,\Omega_S) = \frac{1}{\sigma^{\mathrm{Elastic}}_{pp}}\frac{d\sigma^{\mathrm{Elastic}}_{pp\rightarrow ppS}}{d\Omega_{S}dE_{S}}
\ee
The plot in Fig. (\ref{fig2}) compares the two different methods of calculating the splitting probability as a function of the scalar energy $E_S$. Similar results hold for other choices of $m_S$.
We observe that the ratio is ${\cal O}(1)$, with one pion exchange providing a rate that is slightly larger than the relativistic splitting function for LSND beam energies. This comparison nonetheless provides confidence in the rate calculation at the ${\cal O}(1)$ level.

{\it Scalar production rate:-} Utilizing only the splitting function calculation (\ref{Brem}) as a conservative approximation for the total rate, the total number of scalars $N_S$ produced through the bremsstrahlung channel can be estimated numerically, where we normalize the rate to the number of $\pi^+$ produced, $N_\pi$, which is given at LSND energies by the Burman-Smith distribution \cite{Burman:1989ds}. For $m_{S}=1$ MeV, we obtain $N_{S} \sim 0.5\theta^{2} g_{SNN}^{2}N_\pi$. This calculational approach should capture part of the primary production channel, but as is apparent from the discussion above, it is only anticipated to be accurate up to ${\cal O}(1)$ factors.

\subsection{Other production channels}
In this subsection, we comment on a number of additional sub-leading scalar production channels. 

{\it $\Delta$ decay:-} At LSND beam energies, roughly half the total proton-proton scattering cross section involves an inelastic process with resonant production, e.g. of $\De^{++}$, which subsequently decays to $p+\pi^+$. Indeed, the resonant excitation of $\De$ (and $\Si$) hadronic resonances is the primary channel for pion production at LSND. This is partially incorporated in the analysis of bremsstrahlung above, in that it contributes to the total cross section, but there are additional channels involving $S$ radiation from final states which are more probematic to calculate. A tractable contribution of this type involves 3-body $\De$ decay, $\Delta \rightarrow \pi + p + S$ \cite{Pospelov:2017kep}, as shown in Fig.~\ref{Feynman}(c). Computing the 3-body decay rate, using a phenomenological pion-Delta-nucleon vertex at the low-energy given by $\mathcal{L}_{int} = g_{\pi \Delta N} \bar\De^{\mu} N\partial_{\mu}\pi$, and assuming that the 2{-}body decay of $\Delta$'s saturates pion production inside the target, we can estimate the number of scalars from the 3{-}body decay via the following ratio, $N_{S} \sim N_{\pi}\times \frac{\Gamma_{\Delta \rightarrow p \pi S}}{\Gamma_{\Delta \rightarrow p \pi}}$. Evaluating the 3-body phase space integral numerically for $m_{S}=1$ MeV, we find $N_{S} \sim 0.04 \theta^2 g_{SNN}^{2}N_\pi $, which is consistent with the estimate in \cite{Pospelov:2017kep} and about an order of magnitude below the bremsstrahlung rate. Note that in the collinear limit, scalars are produced isotropically in the $\Delta$ rest frame. We have transformed the energy-angle distribution to the lab frame, using a Monte Carlo simulation, in which the energy-angle distribution of $\Delta$ baryons in the lab frame was reconstructed from the Burman-Smith parameterization of the pion distribution. As expected this distribution is almost isotropic, reflecting the fact that the $\De$'s are produced almost at rest, in comparison to the more forwarded-peaked distribution from bremsstrahlung. This further suppresses the event rate in the detector.

{\it Primakov conversion:-} There are several additional decay channels that will contribute to $S$ production, as discussed in \cite{Pospelov:2017kep}, but none are estimated to be larger than the $\De$-decay channel discussed above. We have also considered a different topology that utilizes the effective diphoton coupling (\ref{diphoton}), via which scalars can be produced via the Primakov conversion of photons $\gamma{+}N{\rightarrow}S{+}N$ in the presence of nuclei with atomic (number) mass $Z$($A$), as shown in Fig.~\ref{Feynman}(d). The dominant source of photons is provided by $\pi^{0}$ decays in the target \cite{Dobrich:2019dxc}. The neutral  pion decay length is roughly $0.1 \;\mu m$ at the LSND beam energy, and thus $\pi^{0}\rightarrow \gamma\gamma$ decays can effectively be treated as a distribution of real photons in the target. Applying the analysis of \cite{Aloni:2019ruo} to scalar production, the total cross section  can be written as $\sigma_{S} = \int d k_{\gamma}d \Omega_{\gamma} f_{\gamma}(k_{\gamma},\Omega_{\gamma}) {\times} \sigma_{\gamma N{\rightarrow}N S}(k_{\gamma})$, where $f_{\gamma}(k_{\gamma}, \theta_{\gamma}) dk_{\gamma} d\Omega_{\gamma}$ is the photon energy and angular distribution with angles $(\theta_{\gamma}, \phi_{\gamma})$ defined with respect to the beam direction. The two-body cross section $\gamma + N\rightarrow S + N$ incorporates a Helm form factor \cite{Woods:1954zz}, which is exponentially suppressed for momentum transfer above $200$ MeV once coherence is lost \cite{Dobrich:2015jyk}. Taking the Burman-Smith model of the pion distributions as an input, the dependence of the photon distribution on the energy and angle with respect to the beam axis was determined using a Monte Carlo simulation, and numerically evaluating the integrals for $m_{S}{=}1$ MeV, we estimate the number of scalars produced as $N_{S}{\sim} 10^{-4} \theta^{2} g_{SNN}^{2}N_\pi$ where the factor of $g_{SNN}^{2}\sim 10^{-6}$ has been inserted purely for comparison. This $S$-production process is forward-peaked, but is subleading at LSND.

\section{Sensitivity at LSND}
\label{sec:sens}

In this section, we focus on the dominant proton bremsstrahlung production mode and combine the rate and distribution of the last section with the experimental geometry and detection probability, in order to determine the LSND constraints on the Higgs portal. The LSND detector was a shielded 5.7m diameter cylinder of length 8.3m filled with 167 tons of mineral oil, that was on average at an angle of 14 degrees to the beamline, and at a distance of 30m from the target. Charged particles, such as electrons and muons, were detected via a combination of Cerenkov and scintillation light. 

Once produced, the probability that an $S$ particle decays inside the detector is
\be
 P_{\rm decay} = e^{-L_i/\gamma \beta \tau} - e^{-L_f/\gamma \beta \tau},
\ee
where $L_i$ (and $L_f$) denote the distances from production at which the scalar will enter (and exit) the detector, while $\tau$ is the lifetime and $\beta$ the velocity. This probability therefore depends on the scalar's energy as well as its direction with respect to beam axis. Due to the low beam energy at LSND, we do not consider scattering or Compton absorption signatures inside the detector, since the decay reach dominates the scattering reach by several orders of magnitude \cite{Pospelov:2017kep, Izaguirre:2017bqb}.

To normalize the overall event rate at LSND, we have used $N_{\pi^{0}}$, the total number of neutral pions produced. In practice, the $\pi^{0}$ distribution is taken to be an average of the measured $\pi^{+}$ and $\pi^-$ production rates in proton-nucleon collisions, which differ by $\mathcal{O}(1)$ factors. For the LSND beam energy, we use the parameterization of the production cross-section given by Burman and Smith \cite{Burman:1989ds}, and denote the total cross section as $\si_\pi^{\rm BS}$. With this normalization, the number of scalars produced via proton bremsstrahlung, that subsequently  deposit their energy in the LSND detector, can be schematically represented as follows,
\begin{align}
N^{\rm LSND}_{S} &\sim \varepsilon_{\rm eff} \frac{N_{\pi}}{\sigma_{\pi}^{\rm BS}}\int dE_{S}d\theta_{S} \bigg( \frac{d^{2}\sigma_{pp_{t}\rightarrow SX}}{dE_{S}d\theta_{S}}\bigg)\nonumber\\
   &\times P_{\rm decay}\vartheta(E_{S}{,}\theta_{S}) 
\end{align}
where $\vartheta(E_{S}{,}\theta_{S})$ summarizes the experimental cut conditions and $\varepsilon_{\rm eff}$ is the corresponding detection efficiency. To determine the sensitivity to scalar decays to electrons, we use the analysis \cite{Athanassopoulos:1997er, Aguilar:2001ty}, in which $\nu_{e}$ were detected via the inclusive charged-current reaction $\nu_{e}{+}\prescript{12}{}C{\rightarrow}e^{-}{+}X$. Following \cite{Essig:2010gu}, we make the assumption, based on the primary use of the scintillation to Cerenkov light ratio, that the $e^{+}e^{-}$ pairs would be registered as indistinguishable from single electrons. Therefore, we assume that the scalar's energy would have been measured as the energy of a single-electron in the energy range $60$ MeV to $200$ MeV with the $e^{+}e^{-}$ pair detection efficiency as for a single electron, \textit{i.e.} $\varepsilon_{\rm eff} \sim 0.1$. A similiar analysis \cite{Auerbach:2003fz} uses an energy cut between $160$ MeV and $600$ MeV on muons produced through the reactions $\nu_{\mu}(\bar{\nu}_{\mu}){+}\prescript{12}{}C{\rightarrow}\mu^{-(+)}{+}p(n){+}X$ in order to identify muon neutrino-like beam excess events inside the detector. We can use this analysis to find the sensitivity to $S$ decays to muon pairs, although the efficiency is harder to estimate in this case given that the $\mu^+\mu^-$ pair will have a lower boost than the corresponding electron decay. We will assume these events are also reconstructed as single-muon events with efficiency $\sim 0.1$ similar to the electron case, but show the results with hatching to indicate that the detection assumptions are distinct. In this case, we also account for the reduced branching fraction to muons when $m_S > 2m_\pi$. In both analyses, the number of beam-excess events does not exceed 20, which we take as the limit for both electron and muon decay channels.

\begin{figure}[t]
\vspace*{-0.0cm}
 \centerline{\hspace*{0.7cm}
 \includegraphics[width=0.52\textwidth]{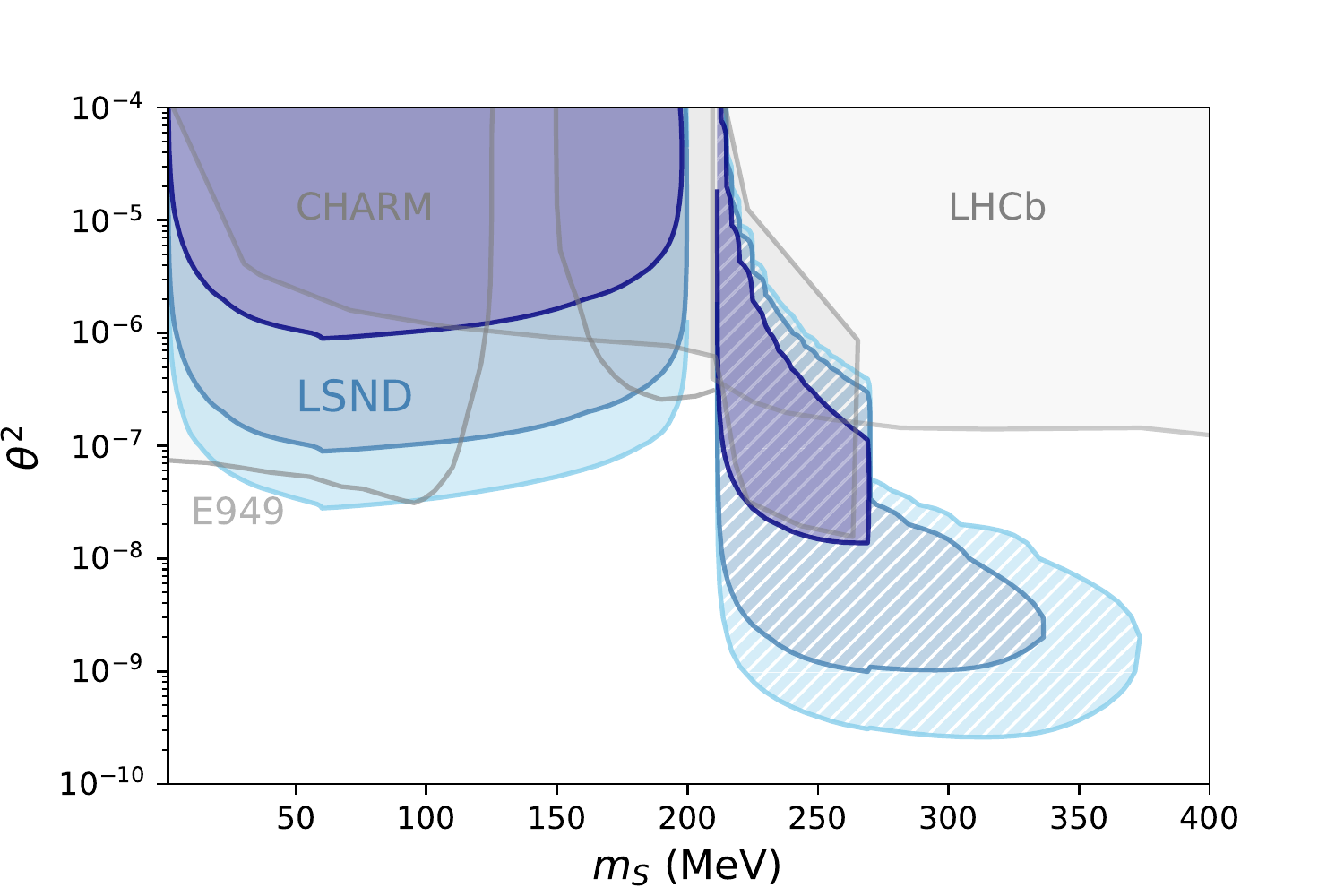}  
 }
 \caption{\footnotesize Sensitivity contours for scalar decays at LSND, with the three blue-shaded contour regions corresponding to 1 event (light), 10 events (medium) and 1000 events (dark). Solid shading indicates event rates from electron decays, while hatched shading indicates event rates from muon decays. Existing exclusions from other sources (in gray) include LHCb \cite{Aaij:2016qsm}, E949 $ K \rightarrow \pi+ invisible$ \cite{Artamonov:2009sz,Clarke:2013aya,Winkler:2018qyg}, and CHARM $S\rightarrow e^{+}e^{-},\; \mu^{+}\mu^{-}$ \cite{Bergsma:1985qz,Clarke:2013aya,Winkler:2018qyg} analyses.}
 \label{sensitivity_contours}
\end{figure}

Using the energy-angle distribution of scalars produced dominantly through proton bremsstrahlung, as outlined in Sec. \ref{production}, and considering the geometric acceptance of the LSND detector as well as kinematic cuts and detection efficiencies for the final state particles, we numerically determined the event yields at LSND. The resulting event number contours are shown in Fig.~\ref{sensitivity_contours}, while our final 20 event limit contour is shown in Fig.~\ref{sensitivity}, which also summarizes the results in comparison to a number of existing constraints as detailed in the Figure caption. We see that the LSND sensitivity to electron decays provides the leading constraint in a small window in scalar mass from 120 to 180 MeV, while the sensitivity to muon decays provides the leading constraint from $2m_\mu$ up to 320 MeV.

\section{Concluding Remarks}
\label{sec:outlook}

In this paper, we have revisited the existing limits on one of the three UV-complete portals from the SM to a dark sector, namely the Higgs portal coupling to a singlet scalar. This portal is of particular interest as one of the generic mediation channels for the interaction with dark matter. We have shown that existing data from LSND, when combined with the dominant low energy production mode through proton bremsstrahlung, already excludes additional regions of parameter space for $m_S$ between 100 and 350 MeV. Future analyses are possible, which can extend this reach further. For example, NA62 at CERN provides greater sensitivity to $K^+ \rightarrow \pi^+ \nu\bar\nu$, and so the exclusion from E949 can be extended \cite{BB3}, while further sensitivity at higher mass may come from Belle II \cite{Filimonova:2019tuy}. Similarly, KOTO provides sensitivity through the neutral decay channel $K_L \rightarrow \pi^0 \nu\bar\nu$ (see e.g. the recent discussions of an anomaly in current data in \cite{Ahn:2018mvc,KOTO,Egana-Ugrinovic:2019wzj,Dev:2019hho,Liu:2020qgx}). The short baseline neutrino (SBN) program at Fermilab will also provide new sensitivity to the Higgs portal, as recently analyzed in \cite{Batell:2019nwo}, and we exhibit the projected sensitivity for SBND and ICARUS from that reference in Fig.~\ref{sensitivity}. 

\section*{Acknowledgements}

We would like to thank D. Karlen, R. Kowalewski, M. Pospelov, R. Tayloe, and K. Tobioka for helpful discussions and communication. This work is supported in part by NSERC, Canada.

\bibliography{Scalar}
\end{document}